\documentclass{article}
\usepackage{graphicx}
\usepackage[a4paper, total={6in, 9in}]{geometry}
\usepackage{float}
\usepackage{hyperref}
\usepackage[dvipsnames]{xcolor}

\title{Super-Resolution Microscopy Based on the Inherent Fluctuations of Dye Molecules}
\author{Alexander Krupinski-Ptaszek$^{1,*}$, Adrian Makowski$^{1,4}$, Aleksandra Mielnicka$^2$, \\Monika Pawłowska$^1$, Ron Tenne$^{3,*}$, 
Radek Łapkiewicz$^{1.*}$}
\date{$^1$Faculty of Physics, University of Warsaw, Pasteura 5, 02-093 Warsaw, Poland
\\$^2$The Nencki Institute of Experimental Biology, PAS, 02-093 Warsaw, Poland
\\$^3$Department of Physics, University of Konstanz, Universitätsstraße 10, D-78457 Konstanz, Germany
\\$^4$Laboratoire Kastler Brossel, ENS-PSL Université, CNRS, Sorbonne Université, Collège de France, 24 rue Lhomond, Paris 75005, France
\\$^*$ \href{mailto:a.krupinski-ptaszek@uw.edu.pl}{a.krupinski-ptaszek@uw.edu.pl}, \href{mailto:ron.tenne@uni-konstanz.de}{ron.tenne@uni-konstanz.de}, \href{mailto:radek.lapkiewicz@fuw.edu.pl}{radek.lapkiewicz@fuw.edu.pl}
}
\begin{document}

\maketitle

\section{Abstract}
Fluorescence microscopy is a critical tool across various disciplines, from materials science to biomedical research, yet it is limited by the diffraction limit of resolution. Advanced super-resolution techniques such as localization microscopy and stimulated-emission-depletion microscopy often demand considerable resources. These methods depend heavily on elaborate sample-staining, complex optical systems, or prolonged acquisition periods, and their application in 3D and multicolor imaging presents significant experimental challenges. In the current work, we provide a complete demonstration of a widely accessible super-resolution imaging approach capable of 3D and multicolor imaging. We replace the confocal pinhole with an array of single-photon avalanche diodes and use the microsecond-scale fluctuations of dye molecules as a contrast mechanism. This contrast is transformed into a super-resolved image using a robust and deterministic algorithm. Our technique utilizes natural fluctuations inherent to organic dyes, thereby it does not require engineering of the blinking statistics. Our robust, versatile super-resolution method opens the way to next-generation multimodal imaging and facilitates on-demand super-resolution within a confocal architecture.

\section{Introduction}
Fluorescence fluctuations in molecules stand at the heart of many super-resolution microscopy (SRM) methods\cite{Betzig2006, Rust2006,Dertinger2009,gustafssonFastLivecellConventional2016,heilemannSubdiffractionResolutionFluorescenceImaging2008,jungmannSinglemoleculeKineticsSuperresolution2010}. In particular, localization microscopy (LM) that regularly achieves 20-30 nm lateral resolutions has made a significant impact on life-science imaging. However, in comparison to confocal microscopy, the tool of choice for life-science imaging, performing LM is a laborious task\cite{vangindertaelIntroductionOpticalSuperresolution2018a}. First, acquisition times can be prohibitively long. Second, three-dimensional (3D) and multi-color imaging remain challenging even two decades after the introduction of LM\cite{lelekSinglemoleculeLocalizationMicroscopy2021}. Finally, successful localization of markers requires a delicate control of the fluctuation statistics of their emission through a precise control of the sample buffer\cite{heilemannSubdiffractionResolutionFluorescenceImaging2008,Dempsey2011,vandelindeDirectStochasticOptical2011} or by controlling the dynamic binding of markers \cite{jungmannSinglemoleculeKineticsSuperresolution2010,jungmannMultiplexed3DCellular2014}. These shortcomings generate a strong drive to find entry-level SRM techniques that maintain the advantages of confocal laser-scanning microscopy (CLSM), even if providing only moderate resolution enhancements. 
One very successful example is structured-illumination microscopy\cite{Gustafsson2000} and its confocal variant image-scanning microscopy (ISM) \cite{Muller2010, Sheppard1988} that has been quickly adopted by commercial systems. In ISM, replacing the confocal pinhole with a small pixelated detector improves the lateral resolution by a factor of two and increases the signal-to-noise ratio (SNR) \cite{York2012, schulzResolutionDoublingFluorescence2013}. A second notable example is super-resolution optical fluctuation imaging (SOFI) in which the temporal correlation of fluctuations forms the contrast for 3D resolution enhancement even without precise engineering of the fluctuation statistics \cite{Dertinger2009, dedeckerWidelyAccessibleMethod2012b, pawlowskaEmbracingUncertaintyEvolution2022}. Recently, we have published a proof-of-principle demonstration of a combination of SOFI and ISM, appropriately termed SOFISM\cite{srodaSOFISMSuperresolutionOptical2020}. Rendering the second-order correlation of the fluorescence intensity for inorganic quantum dots (blinking) provided a x2.5 enhancement of lateral resolution beyond the diffraction limit.

The use of inorganic fluorophores, that served the majority of early SOFI demonstrations, is quite limiting due to the difficulties in incorporating them within cells and targeting specific subcellular sites\cite{Dertinger2009,dedeckerWidelyAccessibleMethod2012b, dertingerAdvancesSuperresolutionOptical2013}. While SOFI images have also been obtained with more common organic fluorophores, such as dye molecules\cite{dertingerSuperresolutionOpticalFluctuation2010} and fluorescent proteins\cite{deschoutComplementarityPALMSOFI2016}, their inherent intensity fluctuations occur at the technologically inconvenient sub-millisecond time scale\cite{s.stennettPhotophysicalProcessesSingle2014}. This scale is incompatible with low-light cameras such as sCMOS, which operate at best with KHz frame-rates and suffer from significant readout noise\cite{schwartzSuperresolutionMicroscopyQuantum2013,VandenEynde2019}.

In recent years, the rapid development of single-photon avalanche diode (SPAD) arrays, in which readout noise is inconsequential, enabled imaging of fluorescence dynamics with sub-nanosecond resolution, six orders of magnitude beyond standard imagers\cite{Bruschini2019,bronziSPADFiguresMerit2016,lubinPhotonCorrelationsSpectroscopy2022a}. Utilizing the nanosecond-scale timing information, fluorescence lifetime imaging (FLIM) and quantum correlations can be measured in widefield (WF) and ISM architectures\cite{Ulku2019, Castello2019, smithVitroVivoNIR2022, Lubin2019, lubinHeraldedSpectroscopyReveals2021, ndaganoQuantumMicroscopyBased2022}. In contrast, the additional capabilities of SPAD arrays at microsecond-to-millisecond time scales were only scarcely used\cite{kloster-landsbergNoteMulticonfocalFluorescence2013, Antolovic2017, Slenders2021}, despite the fact that this is the native timescale for multiple phenomena in single-molecule spectroscopy and biology\cite{krichevskyFluorescenceCorrelationSpectroscopy2002}. 

In the current work, we take advantage of an off-the-shelf confocal SPAD array to capture the native fluctuations of dye molecules and obtain super-resolved SOFISM images of neurological cells. We promote the proof-of-principle demonstration of SOFISM into a fully operational entry-level microscopy technique, performing super-resolution microscopy with standard fluorophores and over reasonably large field-of-views (FOVs) without the use of an imaging buffer. Finally, taking advantage of the pulse-to-pulse resolution of the detector, we use time multiplexing to obtain super-resolved two-color images without the need for image registration in post processing.

\section{Concept}
The excited state of the lowest energy level of a dye molecule is divided into a spin-triplet and a spin-singlet state (Fig. 1a). An absorption of a photon populates the singlet (S=0) state and would be commonly followed by an emission of a photon and a return to the ground state. The inherent fluctuations in the emission intensity result from a transition into the triplet state - intersystem crossing. As radiative relaxation from the triplet to the ground state is dipole forbidden, the excitation is shelved for a period of a few to hundreds of microseconds, depending on the molecule and its environment. Using a standard camera, with a millisecond temporal resolution, the fluctuation dynamics is invisible and manifests only as an undesired decrease of emission yield for labels. 
However, when properly sampled, fluorescence dynamics are also a resource rather than a nuisance for microscopy: they provide extra information that can be applied to overcome the diffraction limit, e.g., in SOFI and LM. 

In our setup (Fig. 1b), the pinhole aperture of a standard confocal microscope is replaced with a confocal 23-pixels SPAD array (SPAD23, Pi imaging). As a fluorescent sample is raster scanned across the laser focus ($\lambda$=635 nm, 40 MHz repetition rate), the detection times of emitted photons are recorded by the SPAD array (Fig. 1c, top). To form images from this dataset, we use two different pipelines: summation and correlation, described in detail in Section 1 of the supplementary information. In the first one, summing the number of photons detected per scan step effectively produces a diffraction-limited confocal laser-scanning microscopy (CLSM) image (Fig. 1d, top right). Alternatively, we bin the detection time of photons with a fine resolution (10 $\mathrm{{\mu}s}$), generating an intensity time trace for each pixel and scan step (Fig. 1c, middle). Calculating the second-order correlation function for every detector pair and summing over a delay range of 50 $\mathrm{{\mu}s}$, we obtain the fast-fluctuations contrast. Performing pixel reassignment for it produces the super-resolved SOFISM image of the Atto643-labeled actin filaments in the sample (Fig. 1d, bottom left) \cite{tenneSuperresolutionEnhancementQuantum2019a,srodaSOFISMSuperresolutionOptical2020}. In pixel reassignment, a concept first introduced in the context of ISM, each pixel effectively acts as an off-optical-axis closed pinhole in a confocal scan. It, therefore, produces an up to x2-resolution-enhanced image that is slightly shifted with respect to the scan position. Combining this SRM method based on pixel reassignment together with that provided by the correlation contrast itself (SOFI), a SOFISM image achieves an up to x4 enhancement of resolution beyond the diffraction limit. To highlight the effect, the two blue frames present magnified portions of the sample analyzed with CLSM and SOFISM. We note that as often done in pixel reassignment, we apply a modified Wiener filter to the SOFSIM data. Details regarding the procedure, termed Fourier reweighting (FR), can be found in the Methods section and Section 1 of the supplementary information.

\begin{figure}[h]
\label{fig1}
\includegraphics[width=\textwidth]{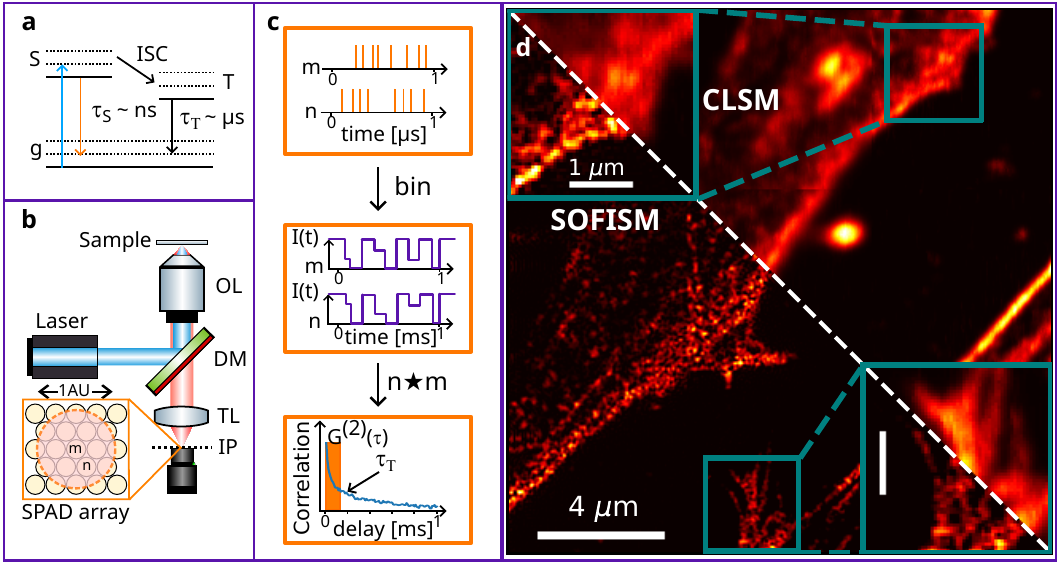}
\caption{\textbf{The concept of SOFISM.} (a) A Jablonski diagram - the energy levels of a dye molecule. After excitation to the spin-singlet state S (blue arrow), inter-system crossing competes with radiative recombination (orange arrow) resulting in the occupation of the spin-triplet state (T). The long lifetime of the dipole-forbidden transition back to the ground state (g) results in a dark period. (b) A schematic of the experimental setup. The inset shows the layout of the 23 pixels in the confocal SPAD array. (c) The SOFISM data analysis pipeline. The bottom panel presents the correlation function average over the entire area depicted in (d), showing an exponential decay with time constants of the order of tens-to-hundreds of microseconds. (d) A demonstration of SOFISM imaging. A neuronal sample, in which actin filaments were stained with Atto643 dye molecules was scanned through the laser focus. The resulting image compares the CLSM analysis (top right) with SOFISM (bottom left). Blue frames present magnifications to better compare the resolution enhancement in SOFISM. To avoid color saturation, the relatively bright bottom-right portion of the image was scaled separately. The scale bar marks 4 $\mu$m in the main image and 1 $\mu$m in the magnified frames. }
\centering
\end{figure}

\section{Results}
For an initial demonstration of the resolution improvement in SOFISM, we performed imaging of a sparse sample of quantum dots (QDs) deposited on glass (see Methods section). The data from a scan of an exemplary scene of two quantum dots separated by approximately 200 nm was processed with CLSM, ISM and SOFISM procedures. To negate the attenuation of high spatial frequencies, we perform FR\cite{Muller2010} on the SOFISM data, yielding an FR-SOFISM image (Fig. 2a). 
The gradual resolution improvement reflects the different contributions to the overall resolution enhancement. The QD pair is unresolved in the CLSM and ISM images, whereas SOFISM clearly separates the two emitters and FR SOFISM further refines the image. A line profile through the centers of the quantum dots is presented in the last panel of Fig. 2a. 
For a quantitative analysis of the resolution enhancement, we analyze 45 measurements of isolated QDs (see Section 2 of the supplementary information). The average full width at half maxima (FWHM) of the point-spread function (PSF) in CLSM, ISM, SOFISM and FR SOFISM is 234 nm, 195 nm, 141 nm and 114 nm, respectively. For comparison, the diffraction-limited performance of the system was analyzed through wide-field imaging of the same QDs, where the average FWHM is 294 nm. Overall, FR SOFISM improves the lateral resolution by a factor of 2.6 beyond the diffraction limit. We note that with the implementation of FR, the resolution enhancement factor is somewhat dependent on the SNR of the contrast through the choice of filter parameters (see Section 1 of the supplementary information). Therefore, for resolution assessment, we use the same parameter values used throughout this paper for image filtering for biological sample scans. 

\subsection{Super-resolved bioimaging with SOFISM}
To verify the straightforward compatibility of SOFISM with life-science imaging, we demonstrate the method for a neuronal-cell sample over a 220 by 220 pixels FOV (Fig. 2). An astrocyte cell in a fixed primary-rat neuronal-cell culture, stained with phalloidin-conjugated Atto 643 dye, is imaged over a 11x11 $\mu\mathrm{m}^2$ FOV. Fig. 2b and Fig. 2c present the images produced by the CLSM and SOFISM analysis pipelines, respectively. Importantly, the staining procedure used here is a common and relatively simple one where the phalloidin peptide binds to the filamentous actin (F-actin) proteins, thus visualizing the cytoskeleton of the cell. 
To highlight the resolution improvement, panels d and e of Fig. 2 present two representative line profiles across the white dashed lines in Fig. 2b. In contrast to the continuous CLSM contour, the correlation contrast exhibits a sharper profile with multiple distinct peaks.
We note that since the labeling density in this sample was insufficient for the achieved resolution of FR SOFISM, the super-resolved images appear somewhat fragmented. Future implementations would require further optimization of the staining procedure.

\begin{figure}[H]
\label{fig2}
\includegraphics[width=\textwidth]{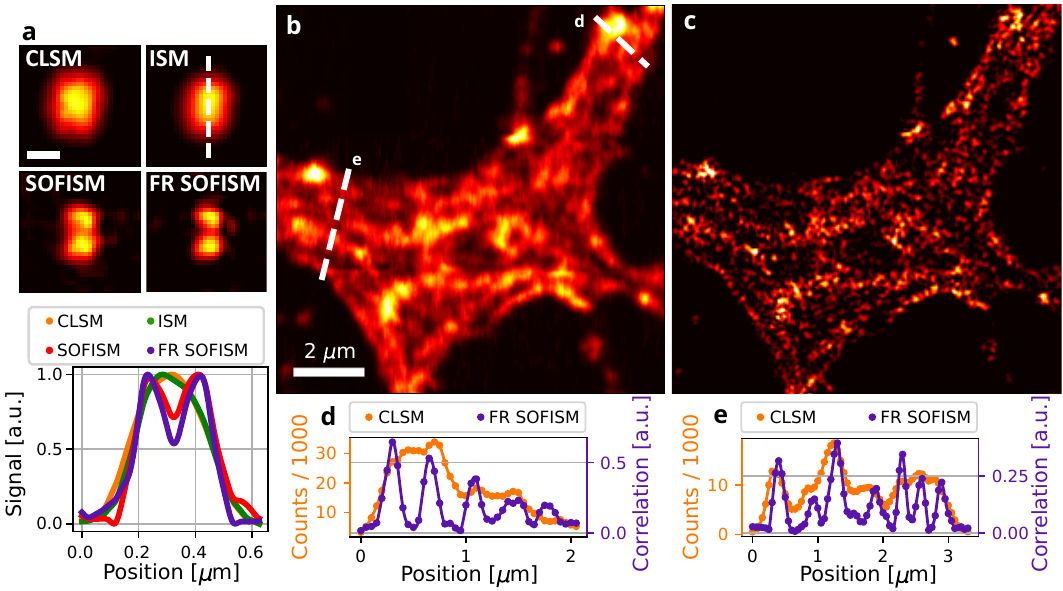}
\caption{\textbf{Analyzing the lateral resolution enhancement of SOFISM.} (a) Comparing the resolving power of CLSM, ISM, SOFISM and FR SOFISM on a scene of two QDs separated by approximately 200 nm (top). Bottom: a comparison of image cross sections taken along the white dashed line. (b,c) A CLSM (b) and FR SOFISM (c) image constructed from the same confocal scan data for an astrocyte cell in which the cytoskeleton is marked with Atto 643 marker molecules. (d,e) Two cross sections taken along the white dashed lines in (c) emphasize the resolution enhancement of SOFISM (purple in comparison with CLSM.}
\centering
\end{figure}
\subsection{Three-dimensional resolution enhancement}
The key benefit of confocal microscopy is optical sectioning - the rejection of out-of-focus contributions to the image, enabling volumetric imaging of biological specimens. Using the correlation contrast, SOFI also introduces sectioning in the axial direction\cite{Dertinger2009,dertingerSOFIbased3DSuperresolution2012}. The blurred emission from a defocused plane mixes the PSFs of multiple emitters and thus reduces the level of single-point temporal correlation. In SOFISM, both effects contribute to a 3D resolution that surpasses that of either methods separately.

To quantify the axial resolution improvement and optical sectioning capabilities of SOFISM, we scan a single QD in three dimensions. CLSM (left) and SOFISM (right) X-Z slices are shown in Fig. 3a (Z is the direction of the optical axis). Clearly, the SOFISM contrast is constrained to a smaller volume, illustrating a 3D resolution enhancement. A white dashed line marks the location of the line profiles presented in Fig. 3b, with FWHM of 642 nm (CLSM, orange) and 389 nm (SOFISM, purple). To quantify the optical sectioning, the intensity and correlation contrast were summed across X and Y, yielding FWHM of 753 nm and 425 nm, respectively. Compared to CLSM, SOFISM improves both the axial resolution of CLSM (x1.65) and the optical sectioning (x1.77) by a similar factor. Additional details regarding the optical sectioning and analysis of volumetric data can be found in Section 3 of the supplementary information.

To verify the improved axial resolution of SOFISM in bioimaging, we perform 3D imaging of the sample shown in Fig. 1 within a 2x2x2 $\mu\mathrm{m}^3$ volume. Fig. 3c presents an X-Y FR SOFISM image of a portion of the cell. A white box marks the region scanned in three dimensions. An X-Z CLSM slice across the white dashed line in Fig. 3c is shown in Fig. 3d. Here, the intensity is spread out across most of the scanned range. A SOFISM analysis from the same scan is shown in Fig. 3e. The much narrower axial confinement of the contrast reflects the improved optical sectioning of SOFISM. A line profile across the Z-axis, matching the white dashed line in Fig. 3d, is shown in Fig. 3f. SOFISM data (purple) depicts fine features that cannot be resolved in the CLSM one (orange). In particular, the resolution improvement is clearly observed by comparing the depth of the dip at Z=0.9 $\mu$m position in Fig. 3f between CLSM and SOFISM. 
Together, the results presented in Figs. 2 and 3 showcase the SOFISM resolution enhancement in all three spatial dimensions using standard dye labels in a biological sample.
In FR SOFISM, the three-dimensional volume of the PSF is x7.21 (FWHM) smaller in comparison with a confocal microscope scan analyzed from the same data set.

\begin{figure}[H]
\label{fig3}
\includegraphics[width=\textwidth]{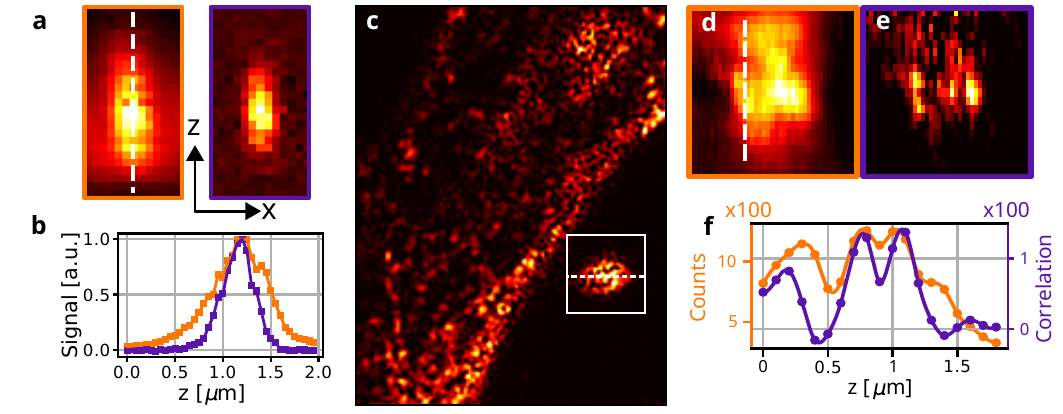}
\caption{\textbf{Axial resolution in SOFISM.} (a) An XZ scan of an isolated QD analyzed through the CLSM (left) and SOFISM (right) pipeline demonstrating a x1.65 axial resolution improvement of SOFISM beyond a confocal microscope. (b) Cross sections along the optical axis from the CLSM (orange) and SOFISM (purple) scans in (a). (c) An FR SOFISM image of a section from the same neuronal sample shown in Fig. 1. The white frame highlights a region in which an XYZ scan was performed. An XZ section of this 3D scan (dashed line in (c)) was analyzed through the CLSM (d) and SOFISM (e) protocols. (f) A Z cross section (dashed line in (d)) demonstrating the improved axial resolution of SOFISM as compared to CLSM (orange). 
}
\centering
\end{figure}
\subsection{Simultaneous multi-color imaging}

Often, understanding biological structure and its relation to function relies on multi-color imaging in which two or more elements are labeled with different fluorophores\cite{batesMulticolorSuperResolutionImaging2007,vangindertaelIntroductionOpticalSuperresolution2018a}. Typically, this is achieved either by sequential snapshots or by spectrally separating the emitted light. In either case, a challenge arises for super-resolved quantitative interpretation of the distances in the form of image registration\cite{erdelyiCorrectingChromaticOffset2013}. The need to precisely align the images from the different channels is often met only by inserting scattering fiducial markers that can be observed in all channels\cite{pertsinidisSubnanometreSinglemoleculeLocalization2010}. In the following, we describe an implementation of two-color super-resolved imaging by temporal multiplexing of the excitation laser pulses. The temporal resolution of the SPAD array sensor is harnessed to streamline two-color super-resolution microscopy and avoid image-registration artifacts.

In our approach, multi-color imaging is achieved by means of pulsed-interleaved excitation\cite{mullerPulsedInterleavedExcitation2005}. Pulse trains from two lasers are synchronized with a delay of $\tau_{C}$ (Fig. 4a). Thanks to the narrow absorption bands of dye molecules, each of the lasers exclusively excites only one type of fluorescent dye. The fluorescence excited by both lasers is then imaged through the same optical path onto a single SPAD array detector without the need to optically divide their contributions. Instead, the channels are easily separated according to the time of detection, i.e. the time delay between the laser trigger and the photon detection. 
The histogram of measured arrival times (Fig. 4b) shows two decaying exponentials with nanosecond-scale lifetimes, distinguished by their delay time from the laser trigger. Emission from AlexaFluor 488 gives rise to the first exponential (green) whereas Atto 647N emission (red) to the second. The separation of the two temporal peaks is controlled by the delay between the two excitation laser pulses (14 ns). A minimal overlap (channel cross talk) is ensured thanks to the short fluorescence lifetime of both dye molecules relative to the delay between the laser pulses. 

Two-color imaging is demonstrated on a primary rat neuronal cell culture where microtubule-associated protein 2 (MAP2) is stained with AlexaFluor 488 (green) and vesicle-associated membrane protein 2 (VAMP2) is stained with Atto 647N (red). The result of a confocal scan (ISM) of 50x50 $\mu\mathrm{m}^2$ FOV of the sample is shown in Fig. 4c. MAP2 (green) is associated with the microtubules, outlining the cytoskeleton of the cell whereas the VAMP2 (red) concentrates mostly around the axons and dendrites. Subsequently, 10x10 $\mu\mathrm{m}^2$ regions, marked with white dashed squares, are imaged with a finer spatial resolution and a longer scan-step dwell time (10 ms per scan step). The CLSM images of these regions are presented in Fig. 4d and 4f. Vesicles (red) are typically observed in the perimeter of the main microtubule strand (green), with some signal with a defocused appearance present in the background. In comparison, Fig. 4e and 4g present the results of FR SOFISM analysis of the data presented in Fig. 4d and 4f, respectively. Co-localization of the two channels is distinctly enhanced in the SOFISM images. In addition, thanks to the improved optical sectioning, SOFISM discards the defocused background contributions of the signals.

\begin{figure}[H]
\label{fig4}
\includegraphics[width=\textwidth]{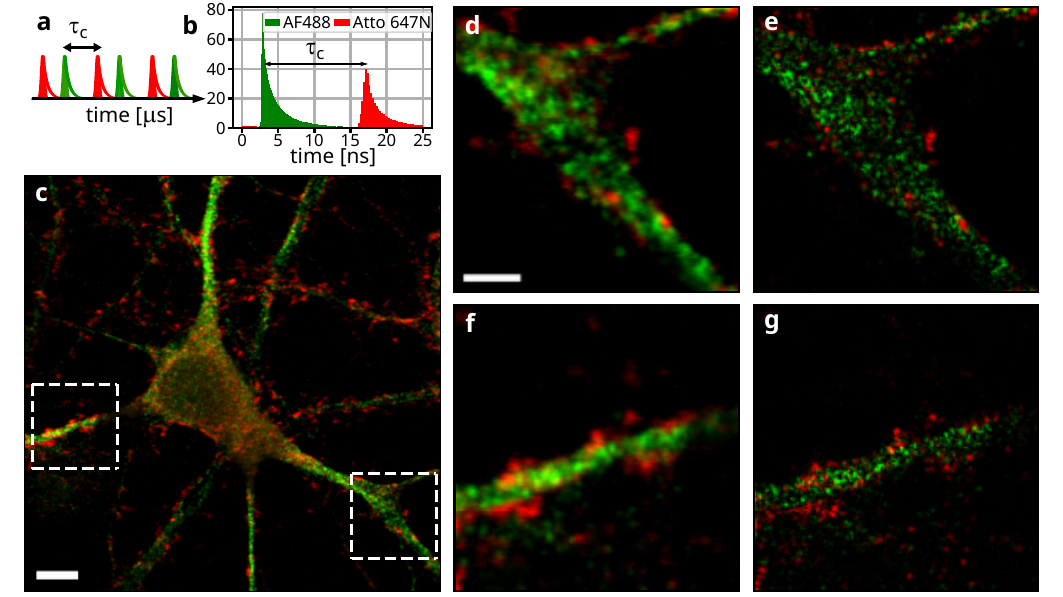}
\caption{\textbf{Implementation of two-color super-resolution with SOFISM.} \textbf{(a)} Scheme of temporal multiplexing. Two laser pulse trains with an interpulse duration of 25 ns each are delayed by 14 ns with respect to each other. \textbf{(b)} The histogram of time delays between the trigger pulses and detection events includes two distinct peaks colored with red and green corresponding to the PL of Atto647N and AF488 molecules, respectively. \textbf{(c)} Two overlaid ISM images from the different channels acquired in a fast scan of a large area. The red and green colors follow the same color scheme of (a) and (b). The two $10 {\mu}\mathrm{m} \times 10 {\mu}\mathrm{m}$ areas marked with white dashed frames are scanned with 10 ms integration time per pixel from which CLSM images (\textbf{(d)} and \textbf{(f)}) and FR SOFISM images (\textbf{(e)} and \textbf{(g)}) are obtained. Scale bars: (c): 5 $\mu$m, (d): 2 $\mu$m. }
\centering
\end{figure}
 To our knowledge, this is the first demonstration of simultaneous multi-color ISM using pulsed interleaved excitation (PIE-ISM). The solution naturally extends to more channels and complements other multi-color strategies, such as those based on PSF engineering or species separation by fluorescence lifetime. 

\section{Conclusions}
In summary, we demonstrated robust 3D and two-color super-resolution bioimaging over a large field of view in a confocal architecture. In our approach, the inherent microsecond-scale fluctuations in the emission of dye molecules are captured by a small SPAD-array sensor that replaces the confocal pinhole. As such, the time-consuming engineering of fluorescence fluctuations, typical to many SRM methods, is unnecessary and the samples are buffer free. 
A x2.09 and x1.65 lateral and axial resolution enhancements over CLSM are demonstrated over large fields of view for an overall x7.21 decrease of the PSF volume. A straightforward extension to two-color imaging is achieved by taking advantage of the nanosecond temporal resolution of the detector.

The scale of invested experimental resources, both in terms of lab work and equipment are similar to that required for confocal imaging. In fact, since the optical setup and staining techniques are nearly identical, the transition between the two modalities is nearly seamless.
Altogether, we believe that SOFISM can serve well as an entry-level SRM - achieving substantial resolution improvement in three dimensions without demanding great efforts in the preparation of the sample or in a special design of an optical setup. Given that cost-effective and low noise SPAD arrays are already becoming a part of commercial confocal microscope systems, SOFISM presents a formidable opportunity for broader adoption of SRM by life-science researchers. An extension to a multispot design, already successfully implemented for ISM\cite{schulzResolutionDoublingFluorescence2013,ingaramoTwophotonExcitationImproves2014,qinDoublingResolutionConfocal2020}, and the algorithmic merger of ISM and SOFISM together \cite{Rossman2021} will enable 3D super-resolved imaging using the inherent fluctuations of dye molecules at rates close to those offered by current confocal microscopes. 
Combined with extensive SPAD array capabilities enabling methods such as fluorescence lifetime imaging \cite{Castello2019, oleksiievetsSingleMoleculeFluorescenceLifetime2022}, anti-bunching microscopy \cite{tenneSuperresolutionEnhancementQuantum2019a, Lubin2019}, and molecule counting \cite{grussmayer2017}
an exciting opportunity arises to leverage the fluctuation contrast in conjunction with optical\cite{huffNew2DSuperresolution2017,tortaroloFocusImageScanning2022} and photophysical information accessible in the multimodal SRM platform to discover new ways of imaging and sensing. 

\section{Methods}
\textbf{Confocal microscopy.} A custom-built microscopy setup was constructed around a Nikon Eclipse Ti2 body. Two picosecond lasers (LDH-D-C-485 and LDH-P-C-635M, PicoQuant) are overlapped on a dichroic mirror (DM) (P3-405BPM-FC-2, Thorlabs) and coupled to a polarization-maintaining single-mode fiber (P3-405BPM-FC-2, Thorlabs). The light output from the fiber is reflected from quad-edge DM (Di01-R405/488/532/635-25x36, Semrock) and focused onto the sample using an oil-immersion objective lens (CFI Plan Apo Lambda D 100X/1.45-NA, Nikon). The emitted fluorescence is transmitted through the DM and quad-line laser rejection filter (ZET405/488/532/642m, Chroma), and imaged through the tube lens (TL) onto a confocal SPAD array (SPAD23, Pi Imaging) placed in the image plane (IP). Detection events providing the time and pixel number of each photons are digitally transferred to a computer for further analysis and image construction. Additional magnification is used to ensure that the SPAD array occupies approximately one Airy unit. The sample is scanned over a stationary focused excitation beam using a piezo stage (P-545.3R7 PInano with E-727.3RD Digital Multi-Channel Piezo Controller, PhysikInstrumente). Both laser and piezo stage provide input trigger pulses to the SPAD array, marking each laser pulse and scan position coordinates, respectively. \\
\textbf{Temporal multiplexing for two-color imaging.} In these experiments both laser pulses excited the sample with a time delay of 14 ns between them (see Fig. 4a and b). The delay between the time tag of each detection and the preceding laser trigger is used is used to discern which of the two laser pulse trains excited this fluorescent transition event. The earlier (later) peak in the histogram in Fig. 4b correspond to the 485 nm (635 nm) laser pulses which exclusively excite two dye markers.
Detections delayed by -0.7 ns to 13.3 ns from the 485 nm laser pulse are considered as that of the first channel whereas the remaining events are assigned to the second channel.  \\
\textbf{Preparation of primary-dissociated hippocampal neurons and glia co-culture for the staining procedure.}
All sample stainings were performed for 3-week-old primary mixed hippocampal cultures. Cells were fixed in $4\%$ paraformaldehyde (Sigma-Aldrich®, 441244) with $4\%$ sucrose preheated to 37℃ for 10 min followed by 3 washes with PBS with $4\%$ sucrose. After fixation, cells were permeabilized with $0.1\%$ Triton X-100 (Bio-Rad, 1610407) in PBS for 10 min, then rinsed 3 times with PBS. Cells prepared this way were then blocked for 1,5 h in $10\%$ goat serum (Gibco®, 16210064) in PBS to block nonspecific sites. \\
\textbf{Immunostaining with antibodies.}
The cultures were incubated overnight at a temperature of 4℃ in a solution containing a mixture of two primary antibodies: mouse monoclonal anti-MAP2 (Sigma-Aldrich®, M1406) diluted 1:500 in PBS with $2\%$ goat serum and rabbit monoclonal anti-VAMP2 (D601A)(Cell Signaling Technology, 13508) diluted 1:250. After they were rinsed 3 times in PBS, the cultures were incubated 2 h at room temperature in a mixture of two various secondary antibodies: goat anti-mouse-IgG-Atto 647N (Sigma-Aldrich®, 50185) and goat anti-rabbit IgG Alexa Fluor 488 (Invitrogen, A11008) both diluted 1:500 in PBS. Cultures were mounted with Fluoromount-G (Invitrogen, 00495802). \\
\textbf{Phalloidin staining.} 
Subsequently, the fixed cells were stained with Atto 643-conjugated phalloidin (Atto-tec). According to the manufacturer, Atto 643 is related to Atto 647N, a popular dye for super-resolution application, but has even higher photostability (\href{https://www.atto-tec.com/fileadmin/user_upload/Katalog_Flyer_Support/ATTO_643.pdf}{Atto 643}). The staining was performed according to the protocol recommended by the manufacturer (\href{https://www.atto-tec.com/images/ATTO/Procedures/Phalloidin.pdf}{phalloidin staining}). The stock solution was prepared by dissolving the 10 nmol phalloidin in methanol to yield a concentration of 10 $\mu$M. To find the optimal emitter density, we stained the sample using 1:240 dilution of stock solution in phosphate-buffered saline (PBS) (4x more diluted than suggested by the manufacturer). Finally, the cover glasses were mounted with ProLong Glass Antifade Mountant with NucBlue nuclear stain (ThermoFisher).\\
\textbf{Preparation of quantum-dots samples.} Samples containing sparse scenes of quantum dots (QDs) for resolution assessment were prepared through drop-casting. A concentrated solution of QDs (Qdot 605 ITK, Invitrogen) was substantially diluted in toluene. A 50 uL drop was then cast onto a microscope coverslip and left to dry in the air. To mitigate QD clustering, prior to drop casting, the QD solution underwent a 15-minute sonication bath.

\bibliographystyle{unsrt}
\bibliography{BioSOFISM_v06.bib}

\section{Acknowledgments}
The authors thank D. Oron, U. Rossman, L. Beck and M. Tillmann for helpful discussions regarding the work and the paper, Wiktor Szadowiak for hardware support, and M. Stefaniuk for support with the Atto643 staining. 

A.K.P. acknowledges the support within “A platform for fast, label-free imaging, identification and sorting of leukemic cells” project (POIR.04.04.00-00-16ED/18-00) carried out within the TEAM-NET program of the Foundation for Polish Science, co-financed by the European Union under the European Regional Development Fund.
R.T. and A.K.P. acknowledge funding by the Deutsche Forschungsgemeinschaft (DFG, German Research Foundation) Project-ID 425217212-SFB 1432. R.T. thanks the Minerva foundation for their support. 
Ad.M. acknowledge National Science Centre, Poland (2023/49/N/ST7/04195) and Scholarship of French Government - Ph.D. Cotutelle/Codirection.
Ad.M., M.P., R.L. acknowledge the support of the Foundation for Polish Science under the FIRST TEAM project “Spatiotemporal photon correlation measurements for quantum metrology and super-resolution microscopy” co-financed by the European Union under the European Regional Development Fund (POIR.04.04.00-00-3004/17-00), by the National Science Centre, Poland, grant number 2022/47/B/ST7/03465, and Horizon Europe MSCA FLORIN project ID 101086142.

\noindent The authors declare no conflicts of interest.
\section{Data availability} Data underlying the results presented in this paper are not publicly available at this time but may be obtained from the authors upon reasonable request.

\end{document}